\documentclass[showpacs,preprintnumbers,amsmath,amssymb,twocolumn]{revtex4}
\usepackage{graphicx}% Include figure files
\usepackage{dcolumn}% Align table columns on decimal point
\usepackage{bm}% bold math
\usepackage{bbm}
\usepackage{amsmath}

\begin{document}

\title{Instabilities of quadratic band crossing points }
% Force line breaks with \\

\author{Stefan Uebelacker}
\email{uebelacker@physik.rwth-aachen.de}
\author{Carsten Honerkamp}
\affiliation{%
Institute for Theoretical Solid State Physics, RWTH Aachen University, D-52056 Aachen 
\\ and JARA - FIT Fundamentals of Future Information Technology
%This line break forced with \textbackslash\textbackslash
}%

\date{15 November 2011}% It is always \today, today,
             %  but any date may be explicitly specified

\begin{abstract}
Using a functional renormalization group approach, we study interaction-driven instabilities in quadratic band crossing point two-orbital models in two dimensions, extending a previous study of Sun \textit{et al.} [Phys. Rev. Lett. 103, 046811]. The wavevector-dependence of the Bloch eigenvectors of the free Hamiltonian causes interesting instabilities toward spin nematic, quantum anomalous Hall and quantum spin Hall states. In contrast with other known examples of interaction-driven topological insulators, in the system studied here, the quantum spin Hall state occurs at arbitrarily small interaction strength and for rather simple intra- and inter-orbital repulsions.
\end{abstract}

% PACS, the Physics and Astronomy
                             % Classification Scheme.
%\keywords{Suggested keywords}%Use showkeys class option if keyword
                              %display desired
\maketitle

\section{\label{introduction}Introduction}
Topological insulators are a new state of electron matter that have attracted enormous interest recently \cite{TIreviews}. Similar to the integer quantum Hall effect, these states can be distinguished from conventional insulators by topological invariants and robust sub-gap surface states. Topological band insulators cannot be continuously deformed into a topologically trivial state without closing the band gap.

In two spatial dimensions and in the presence of time-reversal symmetry, topological insulators are characterized as quantum spin Hall (QSH) states \cite{TIreviews}. If the component of the electron spin along a certain quantization axis is conserved, the picture of the QSH state is especially simple. Then, the currents in the edge states cause opposite Hall conductivities for spin up and spin down, respectively, which results in quantized spin-Hall transport, with the conductance determined by a topological quantum number per spin obtained from the bulk band structure. This simple QSH state can be viewed as two copies of Haldane's 1988 quantum Hall state without net magnetic field \cite{haldane}, with the two copies being related to each other by time reversal. In the case of broken time-reversal symmetry, e.g., if one leaves out one copy or reverts its orbital part, one obtains a related state, known as quantum anomalous Hall (QAH) state.  This state exhibits a quantized charge Hall conductance at zero external magnetic field (see, e.g. Ref. \cite{QiPRB2006}), as in Haldane's initial proposal for the spinless model. In both, QSH and QAH,  the occurrence of edge states through which the electrons propagate comes as a direct consequence of the topologically nontrivial nature of the bulk electronic band structure \cite{TIreviews}.

Usually, this topological structure of the band structure state is caused by the presence of spin orbit coupling. Initially, Kane and Mele proposed a spin-orbit term on the graphene lattice \cite{kanemele}, but it turned out that  for graphene, the effect would be too weak \cite{yao,fabian}. Based on an insightful theoretical proposal \cite{bernevig}, the QSH state could instead be realized in HgTe/CdTe quantum wells, where hallmarks of the QSH effect could be experimentally observed the first time \cite{konig,TIreviews}. 

All this beautiful physics can be understood in the single-particle picture. One could ask, however, if similar topologically non-trivial states can arise from interactions, through a phase transition at some critical temperature. Following this idea, different simple models were analyzed.  Raghu 	\textit{et al.} \cite {raghu} considered the honeycomb lattice with strong next-nearest neighbor repulsion using mean-field and functional renormalization group (fRG) techniques. Above a nonzero critical interaction strength \cite{raghu}, the quantum-spin Hall state was found as the ground state of the system. This instability implies  a spontaneous breaking of the spin rotational invariance. While this study serves as a proof of principle, it is by no means clear how one can realize the rather peculiar interaction with strong second nearest neighbor repulsion. Another study was performed by Zhang \textit{et al.} \cite{3DTMI} for a three-dimensional lattice model where third-nearest neighbor interactions were essential to stabilize the QSH state.  A promising proposal came from Sun \textit{et al.} \cite{sun} who argued, based on mean-field calculations, that quadratic band crossing point (QBCP) models should host QSH states already at arbitrarily weak interactions. The difference to the honeycomb lattice with its Dirac points is that, for a QBCP, the density of states is nonzero at the crossing point, and, hence already a small interaction suffices to drive an instability. This picture is also supported by a mean-field study of Wen \textit{et al.} \cite{ruegg} who found topologically non-trivial phases already for small interactions in Kagome and decorated honeycomb lattice models with quadratic band crossing points.

In this work we consider certain QBCP models \cite{sun} in two dimensions using fRG methods. This allows us to explore the possibilities for interaction-driven topological states beyond mean-field theory. We furthermore investigate if there can be superconducting states emerging from topologically non-trivial insulators.

We briefly note that these interaction-driven topological insulators are also referred to as {\em topological Mott insulators}. The same term is sometimes used in the study of a slightly different question, which is the interplay between a Kane-Mele spin-orbit term leading to a topological band insulator and the Hubbard onsite interaction leading to a Mott insulator, if sufficiently strong \cite{rachel,hohenadler}. Our paper addresses the interaction-induced generation of a Kane-Mele-type term in absence of significant spin-orbit coupling in the bare Hamiltonian, by interactions that are weaker than those required to drive the system Mott-insulating.  A recent work \cite{varney} addresses the transition between topologically non-trivial and trivial states when nonzero interaction parameters are changed in finite-size systems. Interestingly, the single-particle gap can remain robust across the topological transition, and both sides of the transition are insulators.

\section{Continuum fermion model}
\label{continuum model}
We first consider a spin-$1/2$ model in the continuum in two dimensions, which describes the neighborhood of the QBCP in a quite general way. Following Ref. \cite{sun} let the Hamiltonian be given by 
\begin{equation}
H=H_{\text{free}}+H_{\text{int}},
\end{equation}
where the free part reads as
\begin{eqnarray}
\label{hfree}
H_{\text{free}}&=&\sum_{o,o',s} \int d \mathbf{k}\medspace \psi^\dagger_{o,s}(\mathbf{k}) H^0_{oo',s} (\mathbf{k}) \psi_{o',s}(\mathbf{k}) \nonumber \\
&=& \int d \mathbf{k}\medspace \Psi^{\dagger}(\mathbf{k}) \mathbf{H}^0 (\mathbf{k}) \Psi(\mathbf{k}).
\end{eqnarray}
Here the index $s$ denotes spin and $o=1,2$ denotes two different Fermi fields, which should be identified with the orbital degree of freedom. In the second line we have written the Hamiltonian in matrix notation so that $\Psi=(\psi_{A,\uparrow}, \psi_{B,\uparrow},\psi_{A,\downarrow},\psi_{B,\downarrow})^T$ and $\Psi^\dagger=(\psi^\dagger_{A,\uparrow}, \psi^\dagger_{B,\uparrow},\psi^\dagger_{A,\downarrow},\psi^\dagger_{B,\downarrow})$ combines spin and orbital degrees of freedom. We choose the free part to be of the form 
\begin{equation}
\label{h0}
\mathbf{H}^0(\mathbf{k})= I_S \otimes  \left[d_I (\mathbf{k}) I+d_x (\mathbf{k}) \sigma_x+d_z (\mathbf{k}) \sigma_z \right] \, ,
\end{equation}
where $I_S$ is the unity matrix in spin space, I the unity matrix in orbital space, and $\sigma_x$ and $\sigma_z$ the Pauli-matrices. The integral is over a disk in two-dimensional momentum space. The disk radius, i.e. the ultraviolet (UV) cutoff, just determines the energy window focused on and will not be of any qualitative importance. 
The coefficients are $d_I(\mathbf{k})=t_I(k_x^2+k_y^2)-\mu$, $d_x(\mathbf{k})=2 t_x k_x k_y$ and $d_z(\mathbf{k}) =t_z(k_x^2-k_y^2)$. Usually, we will set $t_I$ to zero and $t_x=t_z=t$ for simplicity.  There is no $t_y$  as this would break time-reversal symmetry or already create a QSH state at the bare level. Except for Sec. \ref{sc}, the chemical potential is set to zero so that the QBCP lies at the Fermi level. Note that we exclude the possibility of spin orbit coupling by restricting the Hamiltonian to unity in spin space.

The free part can be easily written in a diagonal basis. With a proper transformation of the Fermi fields $\gamma_{n,s}(\mathbf{k})=\sum_{o} u_{n,o}(\mathbf{k}) \psi_{o,s}(\mathbf{k})$, where $\gamma_{n,s}$ is the transformed field in band $n$ with spin $s$, the free Hamiltonian becomes 
\begin{equation}
\label{freediag}
\mathbf{H}^0 (\mathbf{k})=I_S \otimes \left[ d_I (\mathbf{k}) I+d_z' (\mathbf{k}) \sigma_z \right]  ,
\end{equation}
with $d_z'=\sqrt{d_x^2+d_z^2}$. The band structure of the model consists of two parabolas, which have a QBCP at the origin (see Fig. \ref{continuumplot}). In our choice $t_x=t_z=t$, the dispersion is rotationally invariant in the plane.
 
The interacting part of the Hamiltonian contains local intra- and interorbital repulsions, 
\begin{widetext}
\begin{eqnarray}
H_{\text{int}} & = &\frac{U}{2} \sum_{o,s\ne s'} \int d\mathbf{k} d\mathbf{k'} d\mathbf{q}\, \psi^\dagger_{o,s}(\mathbf{k}) \psi^\dagger_{o,s'}(\mathbf{k'})\psi_{o,s'}(\mathbf{k'}-\mathbf{q})\psi_{o,s}(\mathbf{k}+\mathbf{q})\notag\\
&&+\frac{U'}{2} \sum_{o \ne o',s,s'} \int d\mathbf{k}d\mathbf{k'} d\mathbf{q} \medspace  \psi^\dagger_{o,s}(\mathbf{k}) \psi^\dagger_{o',s'}(\mathbf{k'})\psi_{o',s'}(\mathbf{k'}-\mathbf{q})\psi_{o,s}(\mathbf{k}+\mathbf{q}) \notag\\
&=& \sum_{o_1,o_2,o_3,o_4 \atop s,s'} \int d\mathbf{k}_1 d\mathbf{k}_2 d\mathbf{k}_3 d\mathbf{k}_4\,
V_{o_1 o_2 o_3 o_4}( \mathbf{k}_1,\mathbf{k}_2,\mathbf{k}_3,\mathbf{k}_4) \psi^\dagger_{o_3,s}(\mathbf{k}_3)\psi^\dagger_{o_4,s'}(\mathbf{k}_4 )\psi_{o_2,s'}(\mathbf{k}_2)\psi_{o_1,s}(\mathbf{k}_1) \, .
\label{H_int}
\end{eqnarray}
\end{widetext}
If we use the basis in which the free Hamiltonian is diagonal, we have to transform the interacting part of the Hamiltonian accordingly:
\begin{align}
&\lefteqn{V_{n_1 n_2 n_3 n_4}( \mathbf{k}_1,\mathbf{k}_2,\mathbf{k}_3,\mathbf{k}_4)} \notag\\
&= \sum_{o_1,o_2,o_3,o_4} V_{o_1 o_2 o_3 o_4}( \mathbf{k}_1,\mathbf{k}_2,\mathbf{k}_3,\mathbf{k}_4) \notag
\end{align}
\begin{equation}
\times \quad u_{n_1 o_1}(\mathbf{k}_1) u_{n_2 o_2}(\mathbf{k}_2) u^*_{n_3 o_3}(\mathbf{k}_3) u^*_{n_4 o_4}(\mathbf{k}_4)  \label{omakeup} 
\end{equation}
so that the local interaction (\ref{H_int}) assumes a pronounced momentum dependence in the new basis. The prefactor consisting of the four $  u_{n o}(\mathbf{k})$s is sometimes referred to as orbital make-up.

As pointed out in Ref. \cite{sun},  the QBCP carries a vortex-like winding of the Bloch eigenvectors (with two components for the amplitudes in the two orbitals) that cannot be made undone continuously. The core of the vortex needs to be a degeneracy point, and hence the QBCP cannot be removed easily. When we think about the stability of the QBCP with respect to interactions, there are essentially two ways to remove the QBCP which we will find to be realized depending on the interaction parameters. The free Hamiltonian has rotational invariance in the $k_x$, $k_y$-plane. One way is now to break this rotational symmetry by splitting the QBCP into two Dirac points. In terms of the eigenvectors this stretches out the winding along a branch cut connecting the two Dirac points with an inversion of the direction of the Bloch eigenvectors. The other way is to open a gap by breaking either time reversal symmetry in a given spin sector. This corresponds to a nonzero $d_y$-term.  
Only when the $d_y$s in  spin-up and spin-down sectors are opposite in sign but of the same magnitude, time-reversal symmetry is still present. This state is then a QSH state. Otherwise, time-reversal symmetry is broken. If $d_y$ is spin-independent, we have a QAH state. 

Next we study the instabilities of a continuum QBCP model and of a related lattice variant in the spin-$1/2$ case by a fRG approach. This extends the mean-field study of Sun \textit{et al.} \cite{sun} of a spinless model. We find that a single QBCP might be the most favorable situation to realize spontaneous QSH instabilities.  Compared to the previous studies neither particular longer-ranged nor rather strong interactions are needed, and no threshold value for the interaction strength needs to be exceeded. 
 %Thus even a simple local density-density-repuslion can lead to instabilities towards formation of long range order. 

%To remove the QBCP, the fermionic system must either break the 4 fold rotational symmetry or open a gap by breaking time reversal symmetry or as in the case of the Quantum Spin Hall phase combined 

\section{Functional Renormalization Group treatment}
%In this work we use a momentum cutoff fRG approach \cite{hsfr} to study the weak coupling instabilities of the QBCP. The method is described in this section.
In this section we describe the momentum cutoff fRG method \cite{hsfr,metzner,rgbook}, which is employed in this work.
We consider the action with fermionic Grassmann fields ${\gamma}_s(p)$ and $\bar{\gamma}_s(p)$,
\begin{align}
S(\gamma, \bar{\gamma})=&\int dp \,  \bar{\gamma}_s(p) Q^\Lambda(p) \gamma(p)\notag\\
&+\frac{T}{2} \sum_{s,s'} \int dp_1  dp_2 dp_3 \sum_{n_4} \, V(p_1,p_2,p_3,n_4)\notag\\& \times \bar{\gamma}_s(p_3) \bar{\gamma}_{s'}(p_4) \gamma_{s'}(p_2) \gamma_s(p_1), 
\end{align}
%\begin{equation}
%S(\gamma, \bar{\gamma})=\sum_{i \omega, n} \int d\mathbf{k} \gamma(\mathbf{k},\omega,n) Q^\Lambda(\mathbf{k},\omega,n) \bar{\gamma}(\mathbf{k},\omega,n)\notag\\
%+\sum_{i \omega, n} \int d\mathbf{k}
%\end{equation}
where we used the combined index $p=(\mathbf{k};\omega; n)$ with the Matsubara frequency $\omega$, the band index $n$ as above and $\int dp= \Omega^{-1} \int d\mathbf{k} \sum_{\omega,n}$. $s= \pm \frac{1}{2} $ is the spin projection. $\Omega$ is the Brillouin zone volume, or the considered volume in momentum space.
In the second term, the interaction, we have used wavevector-frequency conservation, and it is only summed over the fourth band index $n_4$.

The idea of the fRG method is to let the quadratic part of the action depend on a cutoff parameter, so that
\begin{equation}
Q^\Lambda(p)= \left[ C^\Lambda(\mathbf{k},n) \right]^{-1} \,  \left[ - i \omega+\epsilon(\mathbf{k},n) \right]
\end{equation}
%\begin{equation}
%Q^\Lambda(\mathbf{k},\omega,n)=T C^\Lambda(\mathbf{k},n) \frac{1}{i \omega+\epsilon(\mathbf{k},n)}.
%\end{equation}
%\sum_{i \omega, n}
with the dispersion $\epsilon(\mathbf{k},n=1,2)=d_I (\mathbf{k}) \pm d_z' (\mathbf{k})$ according to Eq. (\ref{freediag}). For the cutoff-function $C^\Lambda(\mathbf{k},n)$ we choose
\begin{equation}
C^\Lambda(\mathbf{k},n)=\Theta(|\epsilon(\mathbf{k},n)|-\Lambda).
\end{equation}
$\Theta$ is a step function, which is slightly softened in the numerical implementation. With this choice the modes with energy below $\Lambda$ are cut off, so that the free propagator $G^\Lambda_0(p) = - \left[ Q^\Lambda(p)\right]^{-1}$  is restricted to the high energy degrees of freedom.  At $\Lambda=\Lambda_0$ larger than the band width all perturbative corrections are set to zero, as all internal lines in the corresponding diagrams vanish. Hence the correlations and vertex functions of the theory at $\Lambda_0$ are precisely known. The fRG flow from $\Lambda =\Lambda_0$ down to lower $\Lambda$ provides the change of these functions, when the perturbative corrections are successively added to the system.
At $\Lambda=0$ the full action would be recovered, but in the cases and approximations below, the flow of some correlation or vertex functions, in part corresponding to effective interactions, diverge before at a critical scale $\Lambda_c$. This flow to strong coupling is indicative for a change of the ground state. The basic aim of the fRG flows used here is to find out which classes of vertex functions or effective interactions drive this flow to strong coupling. 
From this we obtain tentative ground state diagrams, as described in the sections on the results. We note that the flows to strong coupling are to large extent physically meaningful artefacts of the approximations employed here. If the flow of the fermionic self-energy was included, a gap would open at the instability scale. This would regularize the divergence at nonzero scale. Unfortunately, such flows into the symmetry-broken or massive sector are by far more complicated when the competition of various instabilities or fermionic gaps is kept (which is our goal here). It gets more feasible when only one type of fermion gap is considered \cite{gapflow}. A recent work by Sinner and Ziegler \cite{sinner} shows how the functional RG in partially bosonized form (this formalism is e.g. reviewed in Ref. \cite{metzner}) can be used to determine the final interacting gaps in mono- and bilayer graphene with Coulomb interactions, when all modes are integrated out.
 
 The fRG flow for the vertex functions can be obtained from an exact flow equation for the generating functional of the one-particle irreducible correlation functions \cite{hsfr,metzner,wetterich}. This generating functional can be expanded in the fields, thus yielding an infinite hierarchy of integro-differential equations for the vertex functions. In order to apply the fRG method to our problem, we have to truncate the infinite system by setting the six-point vertex to zero. We also neglect the flow of the self-energy which will only become large, when the interactions flow to strong coupling. We then arrive at the fRG equations %for the self-energy $\Sigma^\Lambda\left( p \right)$  
for the coupling function $V^\Lambda ( p_1,p_2,p_3,n_4 )$, which in a spin-rotationally symmetric situation completely determines the interaction vertex. This equation reads as
\begin{align}
%\label{dgl_sigma}
%&\frac{d}{d\Lambda}\Sigma^\Lambda \left( p \right) =\int dp' S^g \left( p' \right)  \Bigl[ V^\Lambda ( p,p',p' )
%-2 V^\Lambda ( p,p',p ) \Bigr] \notag\\
\label{dgl_v}
&\frac{d}{d\Lambda}V^\Lambda ( p_1,p_2,p_3,n_4 ) =\tau^\Lambda_{PP}+\tau^\Lambda_{PH,d}+ \tau^\Lambda_{PH,cr}\, ,
\end{align}
with the particle-particle channel
\begin{align}
\label{pp}
\tau^\Lambda_{PP}&( p_1,p_2,p_3,n_4 )=- T\int dp \sum_{n'} V^\Lambda ( p_1,p_2,p,n' )\notag\\
&\times L^\Lambda ( p,q_{PP} ) V^\Lambda ( p,q_{PP},p_3,n_4 ) \, , 
\end{align}
the direct particle-hole channel
\begin{align}
\tau&^\Lambda_{PH,d}( p_1,p_2,p_3,n_4) = -T \int dp \sum_{n'} \Bigl[-2V^\Lambda(p_1,p,p_3,n') \notag\\
&\times L^\Lambda(p,q_{PH,d} ) V^\Lambda(q_{PH,d},p_2,p,n_4) + V^\Lambda(p,p_1,p_3,n') \notag\\
&\times L^\Lambda(p,q_{PH,d} ) V^\Lambda(q_{PH,d},p_2,p,n_4) + V^\Lambda(p_1,p,p_3,n')\notag\\
&\times L^\Lambda(p,q_{PH,d} ) V^\Lambda(p_2,q_{PH,d},p,n_4)    \Bigr] \, , \label{phd} 
\end{align}
and the crossed particle-hole channel
\begin{align}
\label{phcr}
\tau^\Lambda_{PH,cr}&( p_1,p_2,p_3,n_4 )= - T \int dp \sum_{n'} V^\Lambda(p,p_2,p_3,n')  \notag\\ 
&\times L^\Lambda(p,q_{PH,cr} ) V^\Lambda(p_1,q_{PH,cr},p,n_4)\, , 
\end{align}
%\begin{align}
%\label{pp}
%\tau^\Lambda_{PP}&( p_1,p_2,p_3,n_4 )=-\int dp \sum_{n'} V^\Lambda ( p_1,p_2,p,n' )\notag\\
%&L^\Lambda ( k,n; -k+k_1+k_2,n' ) V^\Lambda ( p,-p+p_1+p_2,p_3,n_4 )
%\end{align}
%\begin{align}
%\tau&^\Lambda_{PH,d}( p_1,p_2,p_3,n_4) = \notag\\
%&-\int dp \sum_{n'} \Bigl[-2V^\Lambda(p_1,p,p_3,n') L^\Lambda(k,n;k+k_1-k_3,n') \notag\\
%&V^\Lambda(p+p_1-p_3,p_2,p,n_4) + V^\Lambda(p,p_1,p_3,n') \notag\\
%&L^\Lambda(k,n;k+k_1-k_3,n') V^\Lambda(p+p_1-p_3,p_2,p,n_4) + V^\Lambda(p_1,p,p_3,n')\notag\\
%& L^\Lambda(k,n;k+k_1-k_3,n' ) V^\Lambda(p_2,p+p_1-p_3,p,n_4)    \Bigr]\label{phd} 
%\end{align}
%\begin{align}
%\label{phcr}
%\tau^\Lambda_{PH,cr}&( p_1,p_2,p_3,n_4 )= -\int dp \sum_{n'} V^\Lambda(p,p_2,p_3,n')  \notag\\ 
%&L^\Lambda(k,n;k+k_2-k_3,n' ) V^\Lambda(p_1,p+p_2-p_3,p,n_4) . 
%\end{align}
where $q_{PP}=(-\mathbf{k}+\mathbf{k}_1+\mathbf{k}_2;-w+w_1+w_2;n')$, $q_{Ph,d}=(\mathbf{k}+\mathbf{k}_1-\mathbf{k}_3;w+w_1-w_3;n')$, $q_{Ph,cr}=(\mathbf{k}+\mathbf{k}_2-\mathbf{k}_3;w+w_2-w_3;n')$ are the quantum numbers of the second loop line, $p=(\mathbf{k};w;n)$ are those of the first line.  %The integral in these equations is to be understood as an integration over momentum space and a sum over matsubara frequencies $\omega$ and band index $n$: $\int dp=T \sum_{i \omega, n} \int \frac{d \mathbf{k}}{V_{BZ}}$, where $V_{BZ}$ is the volume of the first Brillouin zone. 
Note that we also have to sum over the band index $n'$ of the second internal line, whereas the momentum and frequency are fixed by conservation. The fourth momentum and frequency in the interaction vertex is fixed by conservation, so that here only the band index  is written.
The internal loop is given by
\begin{equation}
L^\Lambda(p,p')=S^\Lambda(p)G^\Lambda(p')+G^\Lambda(p)S^\Lambda(p')\, ,
\end{equation}
with the single-scale propagator
\begin{equation}
S^\Lambda(p)=G^\Lambda(p) \left( \frac{d}{d\Lambda}Q^\Lambda(p) \right) G^\Lambda(p)\, ,
\end{equation}
which has support only at the scale $\Lambda$, as the $\Lambda$-derivative of the cutoff-function is only nonzero there.
In our  approximation of neglecting self-energy corrections, the full propagator $G^\Lambda(p)$ is identical to the free propagator. 

\begin{figure}
\begin{center}
\includegraphics[scale=.3]{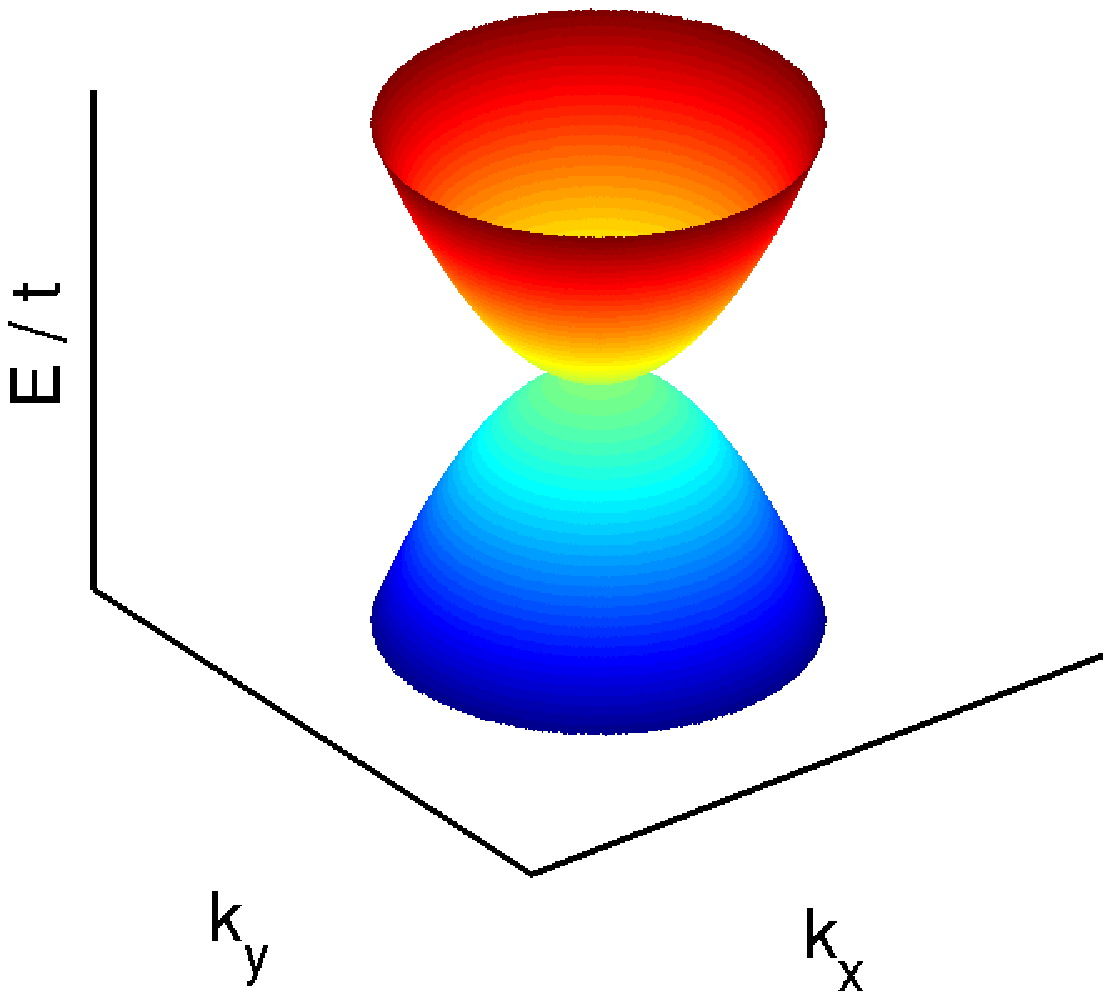}
\includegraphics[scale=.25]{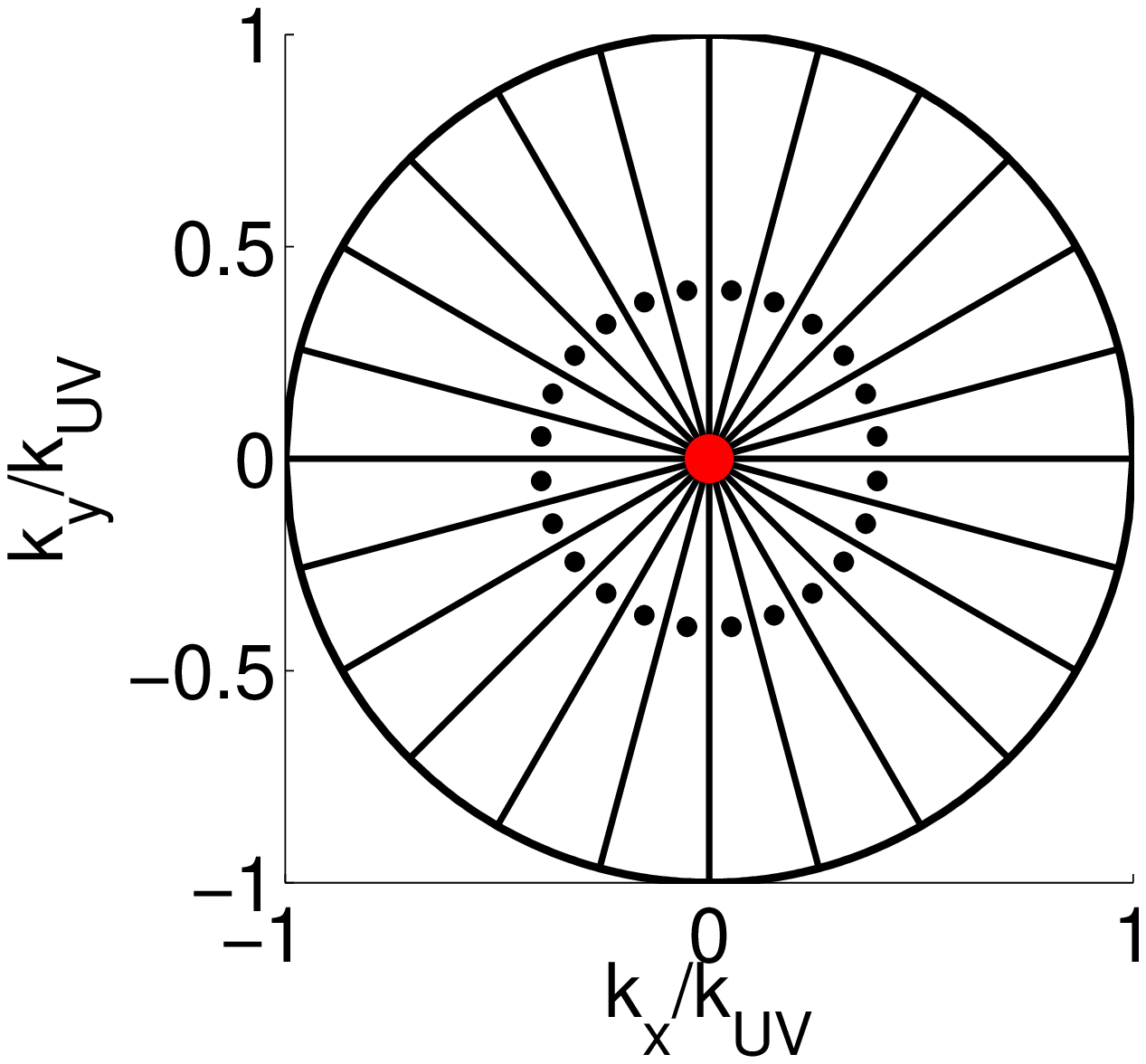}
\end{center}
\caption{Left:  Dispersion relation of the continuum model with the QBCP. Right: Sketch of the patching-scheme: the central red dot is the QBCP, and the black dots denote the momentum vectors $\mathbf{k}$ associated with respective patch, for which the coupling function is computed.}
\label{continuumplot} 
\end{figure}

The frequency dependence of the vertex is ignored in this work, as we are mainly interested in low-frequency vertices that are relevant for static ordering. Thus the frequency summation in the fRG equation can be performed easily analytically. The integration over momentum space is done numerically. For this purpose we divide the momentum space in several patches and compute the coupling function only on a finite set of momenta on a ring around the QBCP, in analogy with previous $N$-patch fRG studies \cite{hsfr,honhon}. Thereby, the coupling function is set constant within each patch (see Fig. \ref{continuumplot}). %This procedure does not correspond to a calculation in a finite system, as the integration of the internal lines  
The advantage of setting the set of momenta on which the coupling function is computed on a ring is, that we can take into account the effect of orbital make up, i.e., the variation and winding of the Bloch functions around the QBCP. The Bloch functions enter the coupling function in band language via the orbital makeup in Eq. (\ref{omakeup}).
This variation can be important for the angular variation of gap structures \cite{honerkamp-epjb} and can certainly change the competition between different instabilities. We note that other RG studies \cite{vafek2010,vafekyang2010} of related problems with two QBCPs contract the wave-vector dependence directly to the Fermi point, i.e. form one big patch and parametrize the interaction by a handful of coupling constants $g$s for the scattering between the Fermi points. While we do not have evidence for a failure of this procedure for the QBCP model, in the iron pnictides, a similar approximation overlooks the angular gap structure \cite{honerkamp-epjb}, and, in principle, the contraction of a winding Bloch function on the center of the winding is ill-defined, at least for a subset of interaction processes. 
Moreover, our more refined patching schemes allows us to study the doped case as well, where the Fermi surface is a small ring around the QBCP point.

We have checked that additional radial patches (i.e., introducing a second ring) do not change the qualitative results stated in this work.
Note that in the continuum the UV cutoff, i.e., the starting scale $\Lambda_0$,  is somewhat arbitrary. Therefore, we choose some fixed UV cutoff. The explicit calculations confirm the expectation that this choice only affects the absolute value of the critical scale, but not the type of instability.

%\begin{align}
%\label{dgl}
%\dot{\Gamma}^{\Lambda} \left( \phi \right) &=\frac{1}{2} Tr \left( C^\Lambda \dot{Q}^\Lambda \right) + \frac{1}{2} \left( \phi, \dot{Q}^\Lambda  \phi \right) \notag\\
%&+ \frac{1}{2} Tr \left( \dot{Q}^\Lambda \left( \frac{\delta^2 \Gamma^\Lambda \left( \phi\right)}{\delta \phi^2}  \right)^{-1}\right) .
%\end{align}

\section{Emergent order parameters and phase diagram in the continuum model}
\begin{figure}
\begin{center}
\includegraphics[scale=0.5]{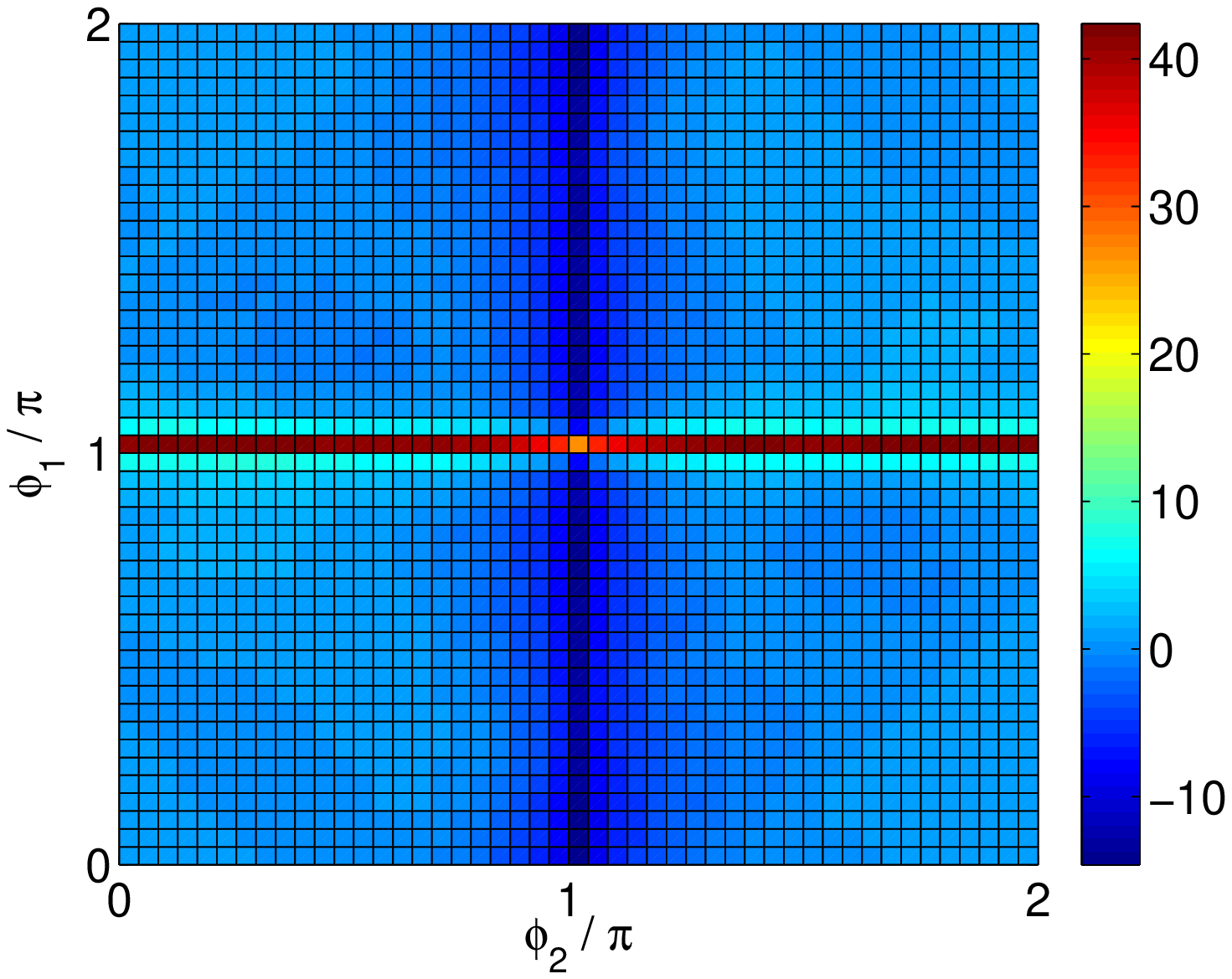}
\includegraphics[scale=0.5]{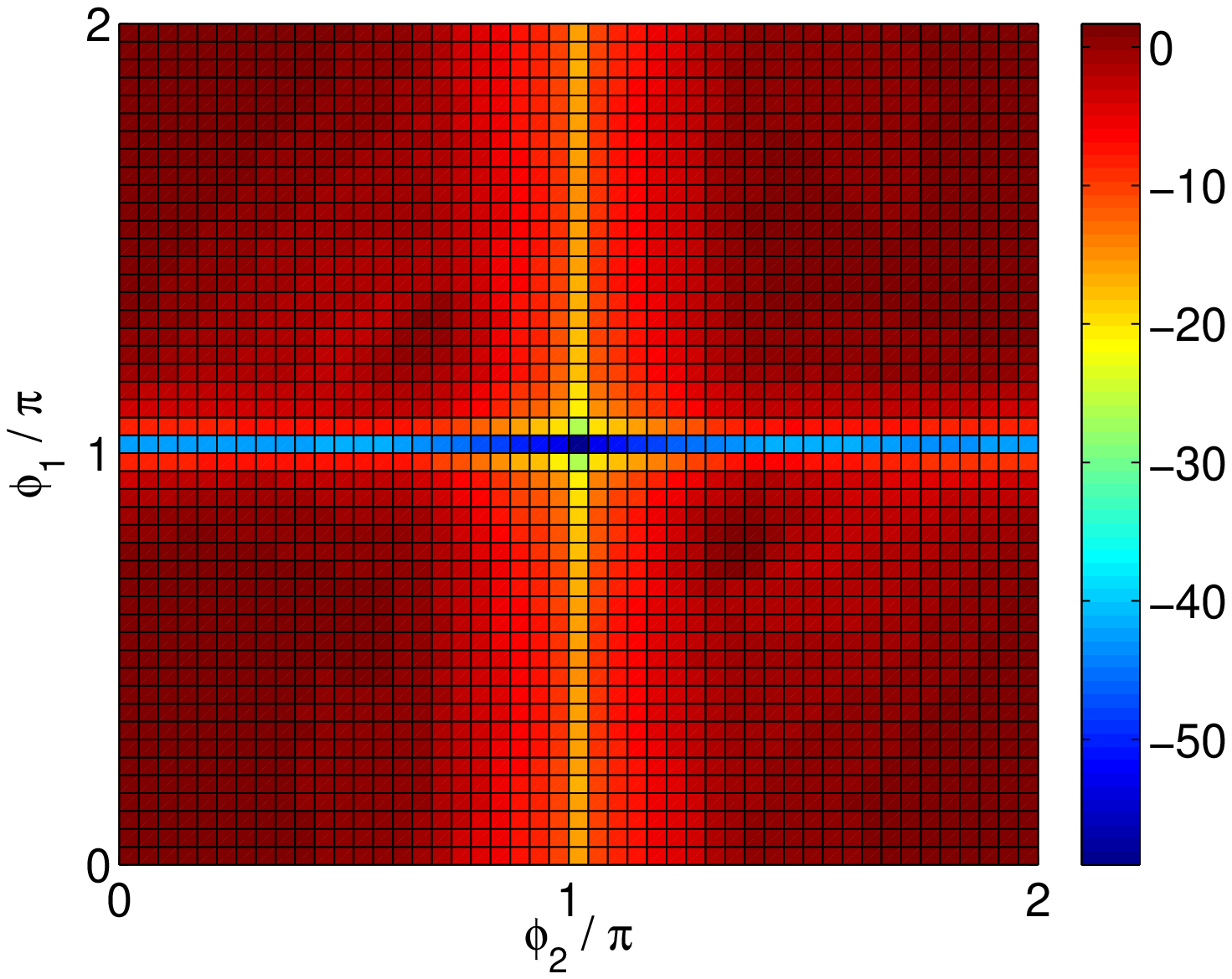}
\end{center}
\caption{Effective interaction near the critical scale in the QAH phase. The angle $\phi_1$ ($\phi_2$) of the first (second) ingoing leg is plotted on the y (x) axis, with the discretization shown in Fig. \ref{continuumplot}. $\phi_3=\pi$ is kept fixed. In our convention the first and the third line have the same spin. The upper plot shows the vertices with the orbitals  $o_1=o_2$, $o_3=o_4 \ne o_1$, the lower plot shows the combination $o_1=o_4$, $o_2=o_3 \ne o_1$. The somewhat weaker vertical features at $\phi_2=\phi_3$  belong to remnants of the SN instability that is predominant at smaller $U'$.}
\label{effww} 
\end{figure}

By investigating the fRG flow of the effective interaction for different ratios of $U$ and $U'$ for the QBCP at the Fermi level we usually encounter flows to strong coupling at some critical scale $\Lambda_c$.  The main growth is in particle-hole channels with zero momentum transfer, leading to very sharp structures in the effective interactions (see Fig. \ref{effww}). If we focus on this dominant zero-momentum transfer component and drop the remaining interaction terms, the effective interaction becomes infinite range in real space and decomposes into a product of two fermion bilinears, each with one creation and one annihilation operator at the same wave-vector. This interaction can then be solved in mean-field theory. Then, the bilinears correspond to order parameters of the strongly coupled phase, the emergence at low scales of which is indicated by the flow to strong coupling. In more detail, we find, depending on the interaction parameters, the following order parameters:
%We investigate the The effective interaction 
\begin{align}
\label{mfnematic}
&\vec{Q}_\text{SN}=\int d\mathbf{k} \langle \Psi^{\dagger}(\mathbf{k})(\vec{\tau} \otimes \sigma_z)\Psi(\mathbf{k})\rangle
  \\ &=\int d\mathbf{k} \sum_{s,s'}\langle \vec{\tau}_{s,s'} [\psi^{\dagger}_{A,s}(\mathbf{k}) \psi_{A,s'}(\mathbf{k})-\psi^{\dagger}_{B,s}(\mathbf{k}) \psi_{B,s'}(\mathbf{k})]\rangle \, ,\notag 
\end{align}
\begin{align}
\label{mfqah}
&\Phi_{\text{QAH}} =\int d\mathbf{k} \langle \Psi^{\dagger}(\mathbf{k})(I \otimes \sigma_y)\Psi(\mathbf{k})\rangle  \\
&=\int d\mathbf{k} \sum_{s} \langle \medspace i \medspace [\psi^{\dagger}_{B,s}(\mathbf{k}) \psi_{A,s}(\mathbf{k}) - \psi^{\dagger}_{A,s}(\mathbf{k}) \psi_{B,s}(\mathbf{k})]\rangle \, ,\notag
\end{align}
\begin{align}
\label{mfqsh}
&\vec{\Phi}_{\text{QSH}} =\int  d\mathbf{k} \langle \Psi^{\dagger}(\mathbf{k})(\vec{\tau} \otimes \sigma_y)\Psi(\mathbf{k})\rangle   \\
&= \hspace{-0.1cm}\int \hspace{-0.1cm}d\mathbf{k} \sum_{s,s'}\langle \vec{\tau}_{s,s'} \medspace i \medspace[\psi^{\dagger}_{B,s}(\mathbf{k}) \psi_{A,s'}(\mathbf{k})-\psi^{\dagger}_{A,s}(\mathbf{k}) \psi_{B,s'}(\mathbf{k})]\rangle \, , \notag
\end{align}
where the brackets $\langle \ldots \rangle$ denote thermal averages.

Here $\vec{\tau} $ are the Pauli matrices in spin space. If we choose $\vec{\tau} $ in the $z$-direction, the state with non-vanishing $\vec{Q}_\text{SN}$ is characterized by a mean-field that is odd both in the two orbitals and in the spin projection along the $z$ axis. For a given spin component, the QBCP splits into two Dirac points, either along the $k_x$ axis or along the $k_y$ axis. For the opposite spin projection, the dispersion is rotated by $90 ^\circ$. The translational symmetry however remains conserved. Therefore this state exhibits spin nematic (SN) order. 

The order parameter $\Phi_{\text{QAH}}$ describes a QAH phase, in which time reversal symmetry is broken. This state has a gapped bulk spectrum, as can be found out from adding the mean-field to the free Hamiltonian. The state then has a quantized Hall conductivity and topologically protected edge states, as can be understood, e.g. by computing the Hall conductivity from the skyrmion-number formula and by looking at finite systems in real space \cite{QiPRB2006}.

The QSH phase corresponds to the third order parameter $\vec{\Phi}_{\text{QSH}}$. This phase is also gapped and breaks spin rotational symmetry. It has helical edge states, which results in quantized spin Hall conductivity.

%\section{fRG results for the continuum model}
Now let us discuss the parameter regions where these orders emerge in the continuum model in more detail. When we run the fRG with the local interaction parameterized by $U$ and $U'$  as initial conditions, we find different effective interactions dominated by the three types of ordering listed above in Eqs. (\ref{mfnematic}), (\ref{mfqah}) and (\ref{mfqsh}) as a function of $U'/U$.
From analyzing which component of these three grows most strongly we can deduce a tentative phase diagram describing the leading ordering tendencies. Here, the critical scale $\Lambda_c$ serves as an (upper) estimate for possible ordering temperatures, or at least for the onset of strong correlations of the type indicated. Note that the question as to whether these orderings actually occur in true long-range form or are prohibited by, e.g., collective fluctuations is not answered in this fRG scheme. Nevertheless, the analysis is expected to give a realistic account of the dominant non-local correlations. Another interesting possibility would be that the single-particle Hamiltonian already contains small terms corresponding to the addressed mean fields, e.g., due to spin-orbit coupling, and that these terms then get strongly enhanced at low temperatures due to the interaction effects monitored by the fRG. 

\begin{figure}
\begin{center}
\includegraphics[scale=0.5]{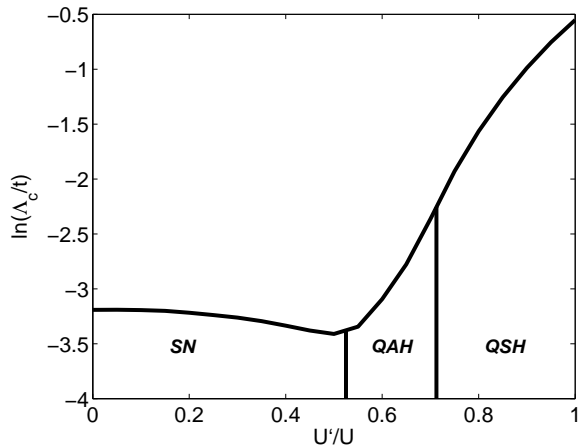}
\end{center}
\caption{Tentative fRG phase diagram of the continuum model as a function of the interorbital interaction parameter $U'$ in units of the intraorbital repulsion. $\Lambda_c$ denotes the critical scale, defined as the scale at which the couplings exceeds three times the bandwidth, obtained with a fRG calculation using $N=48$ patches per band, upper cutoff $\Lambda_{\text{UV}}=10t$ and $U=8t$. }
\label{pd-continuum} 
\end{figure}

The phase diagram obtained for the QBCP model is shown in Fig. \ref{pd-continuum}. For small values of the interorbital repulsion $U'$ we encounter the SN phase with suggested order parameter (\ref{mfnematic}). For $U'\gtrsim U/2$ the QAH-order parameter (\ref{mfqah}) is leading and above  $U'\gtrsim 0.7 U$ the system is unstable toward the QSH-order (\ref{mfqsh}). 

In Fig.  \ref{effww} we show a typical plot of the effective interaction at the critical scale in the QAH phase. One nicely observes the sharp momentum dependence of the $\mathbf{q}=0$-instability in the strong horizontal features. Note also the different sign of the couplings with $o_1=o_2$ and $o_1\ne o_2$ respectively. This reflects the fact that the expectation value of $\psi^{\dagger}_{A,s}(\mathbf{k}) \psi_{B,s}(\mathbf{k})$ is imaginary and thus shows that the corresponding QAH mean-field breaks time reversal symmetry. The less pronounced vertical features correspond to a tendency  toward the emergence of the SN phase, which is still present but weaker than the leading QAH instability.
The other instabilities can also be inferred from analyzing these snapshots of the fRG flow near the instability for different values of the interaction parameters. 

Upon changing $U'/U$, $\Lambda_c$ does not decrease significantly between different phases. This suggests that there is no direct competition between the different tendencies, and the phase transitions are first order.

In order to answer the question as to whether there is a critical interaction strength needed for the emergence of long range order, we also investigated the critical scale in dependence of the bare coupling strength for fixed $U'/U$ in all three regimes. We observed that the couplings do indeed diverge down to an interaction strength of less then $U=2t$ for a bandwidth of $20t$. Although an arbitrary small interaction strength is not numerically accessible, we expect the instability to persist to infinitesimally small interaction.

%The critical scale for fixed $U'/U$ ratio in the QSH regime  goes as $\Lambda_c \propto \exp (-1/(\text{DoS} \times U)$, where DoS is the density of states at $\mathbf{k}=0$. This has been confirmed down to an interaction strength at which $\text{DoS} \times U$ of the order of $1/15$ (or  $U\sim 1/20$ of the bandwidth). Although an arbitrary small interaction strength is not numerically accessible, we expect the instability to persist to infinitesimally small interaction strength.

The conclusion we can draw from this study is that this QBCP model is a favorable situation for the occurrence of spontaneous spin Hall instabilities. No particularly strong or exotic interaction is needed, and the ratio $U'/U \le 1$ between interorbital and intraorbital interactions should not be too unrealistic. The main difficulty regarding possible realizations of these phases is the dispersion with the QBCP. We will come back to this in the conclusions.

Note that the $d$-wave-like wave-vector dependence of the components of the free Hamiltonian and its off-diagonal content that lead to a wave-vector-variation of the Bloch eigenvectors are essential for obtaining these interesting instabilities. If we had just taken a Hamilton matrix $ \mathbf{H}^0(\mathbf{k}) \propto  \mathbf{k}^2 \sigma_z$ that leads to the same dispersion but constant Bloch eigenvectors, the dominant instabilities would have been Stoner ferromagnetism at small $U'/U$ and orbital ordering at larger $U'/U$. These states are neither SN nor support topologically protected edge states. 

\section{QBCP on a Checkerboard Lattice}
\begin{figure}
\begin{center}
\includegraphics[scale=0.45]{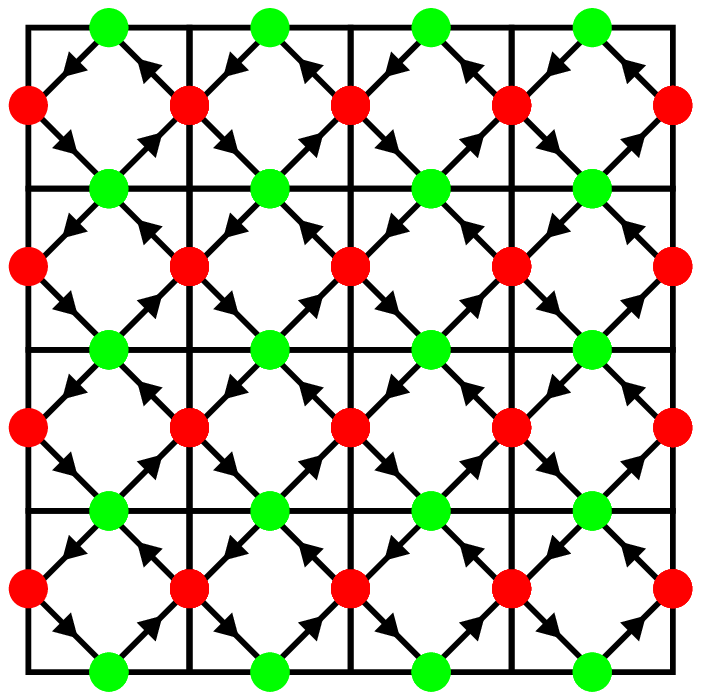}
\hspace{0.2cm}
\includegraphics[scale=0.3]{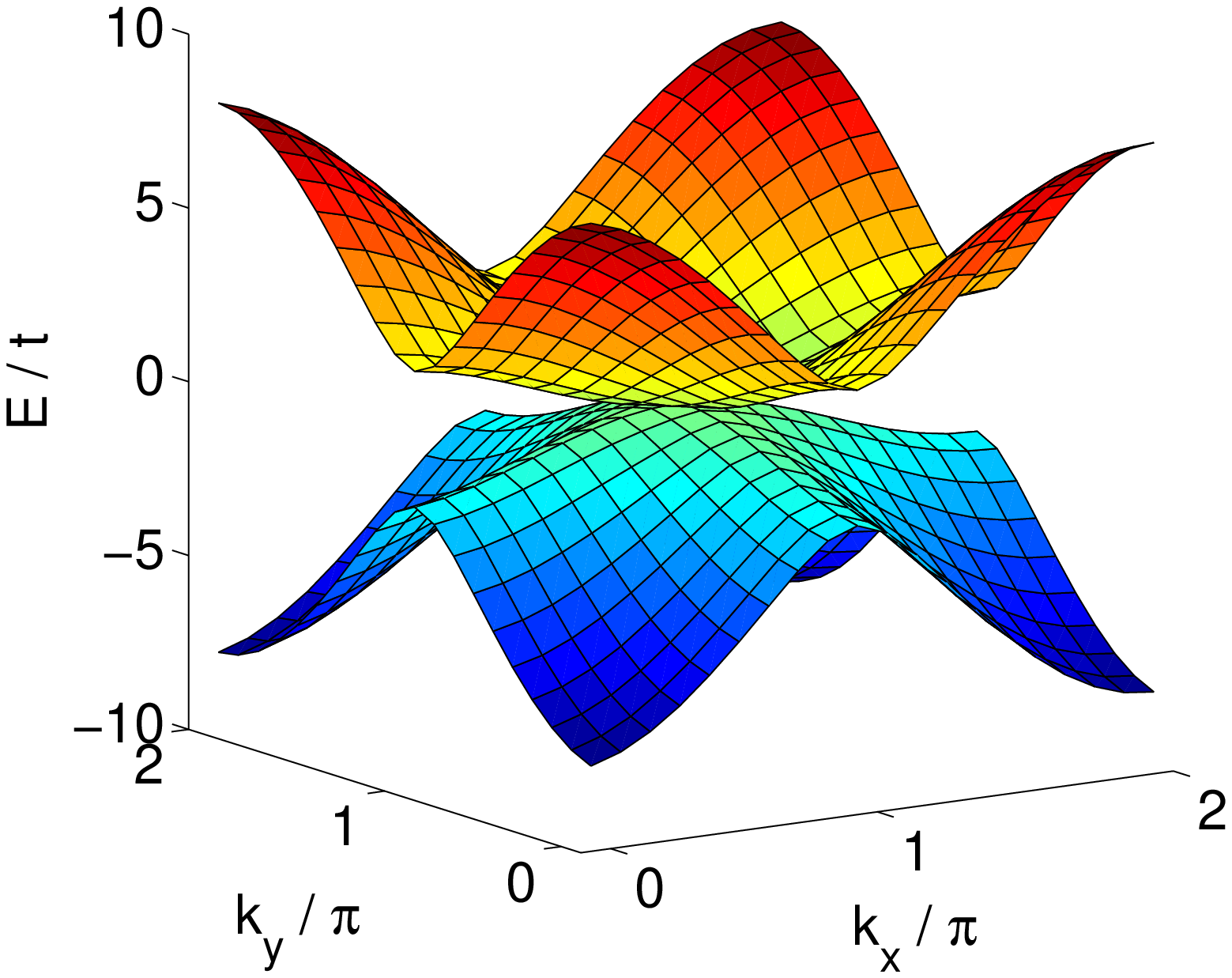}
\end{center}
\caption{Left: Sketch of the Checkerboard lattice with two sublattices A and B indicated by the two different colors. The arrows indicate the current pattern in the QAH and QSH phase. In the QSH phase the current for opposite spins is reversed.  
 Right: Dispersion of the lattice Hamiltonian of Eq. (\ref{latham}). }
\label{lattice} 
\end{figure}

For possible realizations of the unconventional particle-hole instabilities found in the preceding section, it would further be advantageous to have a lattice model that shows similar physics.
In this section we want to analyze a QBCP on a checkerboard lattice, which has two sublattices $A$ and $B$ playing the role of the orbital degree of freedom in the last section. This model was already studied by Sun \textit{et al.} in mean-field theory for spinless fermions \cite{sun}. Here, we treat the spinful case with the fRG.
 The free Hamiltonian is still of the form of Eqs. (\ref{hfree}) and (\ref{h0}) but with the new hopping terms
\begin{align}
d_I&=2 t_I(\cos k_x + \cos k_y )-\mu \, ,\notag\\
d_x&=8 t_x \cos \frac{k_x}{2} \cos \frac{k_y}{2} \, ,\notag\\
d_z&=2 t_z (\cos k_x -\cos k_y) \, .  \label{latham}
\end{align}
Here, the lattice constant has been set to unity.  $t_x$ corresponds to hopping between nearest neighbors and $t_I$ and $t_z$ correspond to hopping between next nearest neighbors, that is nearest neighbors on the same sublattice. If we again set $t_I=0$, then the hoppings between next nearest neighbors connected by a line in Fig. \ref{lattice} and not connected by a line have opposite sign. We also set $t_x=t_z=t$ and $\mu=0$ as in the last section.

%then the hopping in $x$- and $y$-direction on the same sublattice as well as the hopping on different sublattices but with the same direction have opposite sign.
This lattice model has a QBCP around the corner of the Brillouin zone $(\pi,\pi)$ and in its vicinity the lattice model assumes the form of the continuum model in the last section. The dispersion is shown in the right panel of Fig. \ref{lattice}. Further away from the band crossing, the dispersion breaks rotational symmetry around $(\pi,\pi)$.

We consider an onsite repulsion $U$ and a spin-independent nearest-neighbor repulsion, thus, the interaction part of the Hamiltonian reads as
\begin{eqnarray}
H_{\text{int}}\hspace{-0.3cm}&&= \frac{U}{2 N}\hspace{-0.1cm} \sum_{\mathbf{k},\mathbf{k'},\mathbf{q} \atop o,s \ne s'} \medspace \psi^\dagger_{o,s}(\mathbf{k}) \psi^\dagger_{o,s'}(\mathbf{k'})\psi_{o,s'}(\mathbf{k'}-\mathbf{q})\psi_{o,s}(\mathbf{k}+\mathbf{q})\notag\\
&&+\frac{U'}{4 N}  \sum_{\mathbf{k} , \mathbf{k'} , \mathbf{q}  \atop o \ne o',s,s'} \medspace \left(\cos \frac{q_x+q_y}{2} + \cos\ \frac{q_x-q_y}{2} \right) \notag \\ && \times \quad\psi^\dagger_{o,s}(\mathbf{k}) \psi^\dagger_{o',s'}(\mathbf{k'})\psi_{o',s'}(\mathbf{k'}-\mathbf{q})\psi_{o,s}(\mathbf{k}+\mathbf{q}) \, ,
\end{eqnarray}
where we included an additional factor $1/4$ in the second term to compensate that each lattice site has 4 neighboring sites, so that the results are directly comparable to the continuum model.

We carry out a similar fRG-analysis as described in Sec. \ref{continuum model}. Again, for small $U'/U$  we encounter a SN phase which has the same order parameter as in the continuum model given by Eq. (\ref{mfnematic}).
The mean fields of the QAH and QSH phases suggested by the fRG, however, have an additional prefactor due to the local separation of the orbitals and can be written as

\begin{align}
\label{mfqahlattice}
\Phi_{\text{QAH}}&= \sum_{\mathbf{k}} {\Big \langle} \Psi^{\dagger}(\mathbf{k})(I \otimes \sigma_y)\Psi(\mathbf{k}) \Big(-2 \sin \frac{k_x}{2} \sin \frac{k_y}{2} \Big ) {\Big\rangle} \notag\\
&=\frac{1}{2}\sum_{j,\delta,s}\langle \medspace i \medspace D_{\delta}[\psi^{\dagger}_{B,s}(j+\delta) \psi_{A,s}(j) \notag\\&-\psi^{\dagger}_{A,s}(j) \psi_{B,s}(j+\delta)]\rangle \, ,
\end{align}

\begin{align}
\label{mfqshlattice}
\vec{\Phi}_{\text{QSH}}&= \sum_{\mathbf{k}} {\Big\langle} \Psi^{\dagger}(\mathbf{k})(\vec{\tau} \otimes \sigma_y)\Psi(\mathbf{k}) \Big(-2 \sin \frac{k_x}{2} \sin \frac{k_y}{2} \Big){\Big \rangle}\notag\\
&=\frac{1}{2}\sum_{j,\delta,s,s'}\langle \vec{\tau}_{s,s'} \medspace i \medspace D_{\delta}[\psi^{\dagger}_{B,s}(j+\delta) \psi_{A,s'}(j)\notag\\&-\psi^{\dagger}_{A,s}(j) \psi_{B,s'}(j+\delta)]\rangle \, ,
\end{align}

with $j$ denoting the lattice sites of sublattice $A$, $\delta=\pm \frac{x}{2} \pm \frac{y}{2}$, $D_{\pm( \frac{x}{2} + \frac{y}{2})}=1$ and $D_{\pm( \frac{x}{2} - \frac{y}{2})}=-1$. Note that the topological nontrivial phases do not violate local charge conservation despite their unusual appearance. It can be shown that the expectation value of the fermion number operator at a given coordinate remains constant. The current pattern in the QAH state is indicated by the arrows shown in Fig. \ref{lattice}. In the QSH state we get the same pattern for one spin component, but here the current for the other spin component is reversed.

\begin{figure}
\begin{center}
\includegraphics[scale=0.5]{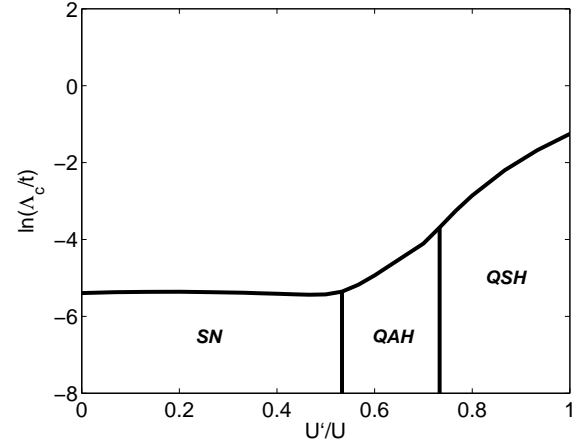}
\end{center}
\caption{Phase diagram of the lattice model as a function of the interorbital interaction parameter $U'$, in units of the intraorbital repulsion $U=6t$. $\Lambda_c$ denotes the critical scale, defined as the scale at which the couplings exceeds four times the bandwidth, obtained with a fRG calculation using $N=48$ patches per band.}
\label{pd-lattice} 
\end{figure}

The phase diagram of the lattice model is shown in Fig. \ref{pd-lattice}. It is similar to the one of the continuum model, both phase transitions occur at similar $U'/U$-ratio. As expected the difference between the two models, which lies only in the high energy modes, do not play an essential role for the determination of the leading instability. The low energy instability of a model with a QBCP at the Fermi level can be reasonably well approximated by the continuum model and is presumably quite independent of details of the band structure in the high energy sector.

\section{Flows at non-vanishing chemical potential}
\label{sc}

We now want to investigate the instabilities of the QBCP at a finite value of the chemical potential so that the QBCP is shifted away from the Fermi surface. We mainly consider the continuum model of Sec. \ref{continuum model}. However we checked that the results remain qualitatively similar in the checkerboard lattice of the last section, though. 
 
The natural choice for the patching-points of the discretization scheme is to set them on the circular Fermi surface that opens upon changing the chemical potential away from zero, as we are interested in the effective model at low energy. We choose the patching points of the other band without Fermi surface to be at the same positions in momentum space. Again we also used a momentum discretization with additional radial patches.

\begin{figure}
\begin{center}
\includegraphics[scale=0.5]{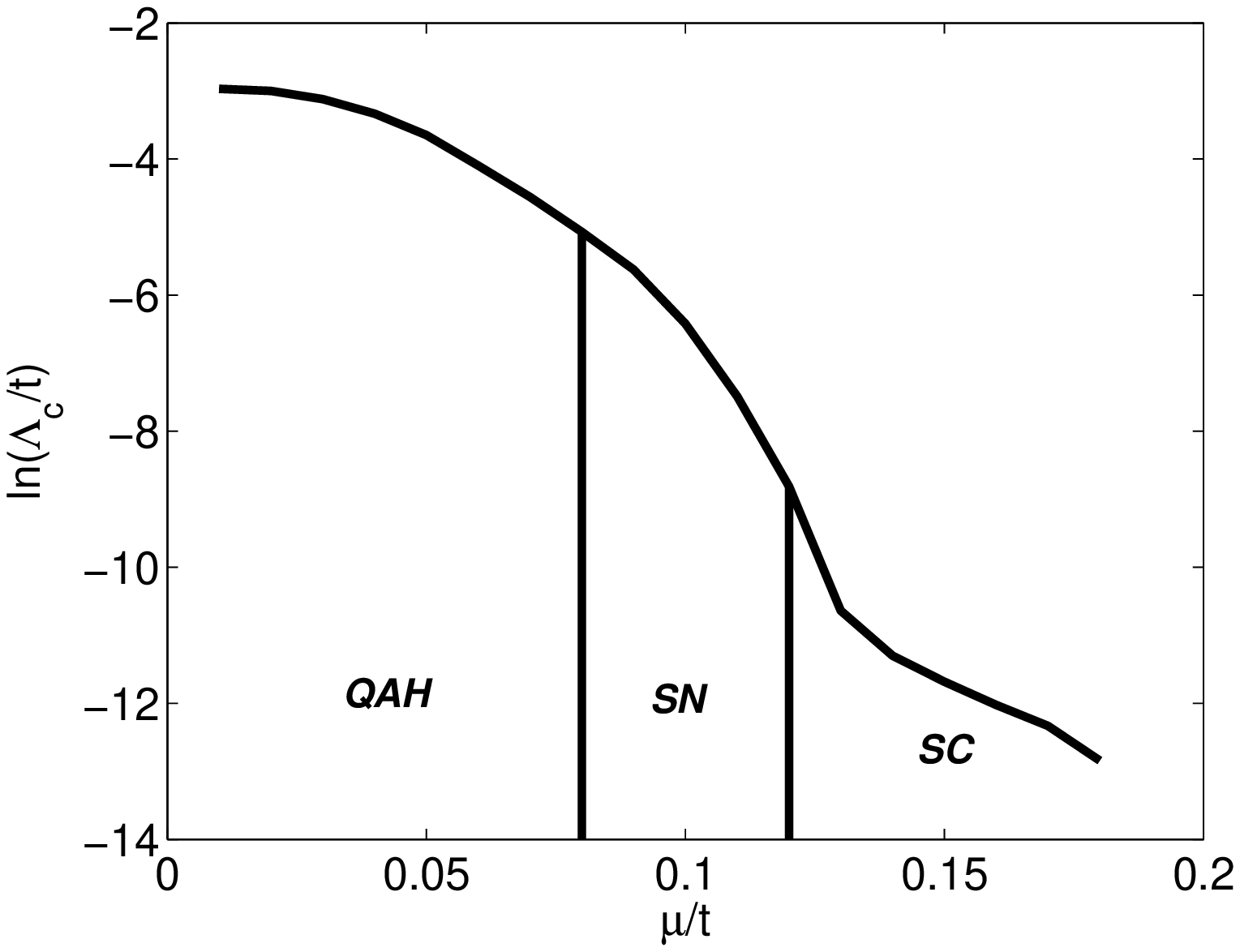}
\includegraphics[scale=0.5]{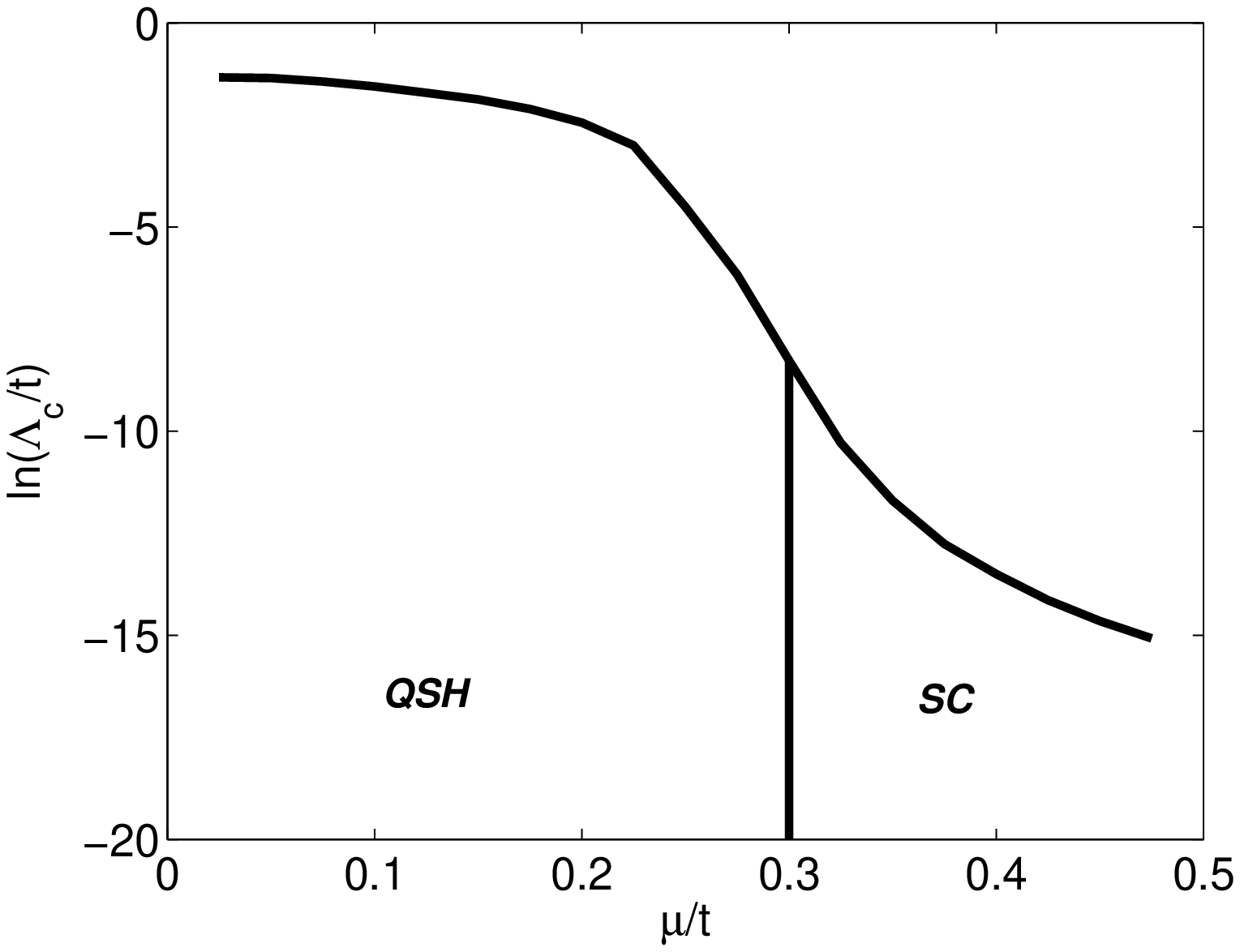}
\end{center}
\caption{Phase diagram of the continuum model as a function of the chemical potential $\mu$ at $U'/U=0.55$ (upper plot) and $U'/U=0.75$ (lower plot)}
\label{pdsc} 
\end{figure}

In all three regimes, we find a critical chemical potential $\mu_{\textrm{crit}}$, above which the leading instability is superconducting. This means that the leading divergence is now for momentum combinations that have total incoming momentum, $\mathbf{k}_1+\mathbf{k}_2=0$ (assuming that the QBCP is at the origin in momentum space). 
%Generally $\mu_{/textrm{crit}}$ increases in the QSH regime with larger $U'/U$-ration, which corresponds to the increase of the critical scale for $\mu=0$.
Typical phase diagrams are shown in Fig. \ref{pdsc}. We see that the critical scale drops for larger $\mu$.

A further interesting result is, that in the regime, where we get the QAH for $\mu=0$, there is always an intermediate phase, either SN or QSH type, before the superconducting channel becomes strongest. The leading instability of the intermediate phase is determined by the ratio of $U'$ and $U$: above $U'/U=0.6$, it is QSH and below it is SN.

We also addressed the question as to which type of pairing symmetry can be associated to the superconducting instability. It turns out that this question can not be conclusively answered within our approach. Depending on numerical details, in particular, the discretization scheme (i.e., use of one or more radial patches), we obtain two different results, which are most plausible according to our data and which both have an orbital dependence of the order parameter.  In the first case, the pairing has an interorbital $s$-wave-like component and a $d$-wave-like intraorbital component. The second, nearly degenerate, possibility is an odd-parity, i.e., triplet pairing, symmetry that is $p$-wave-like in all orbital combinations, which has relative phase shifts between the orbital components.
From our data, it seems that different superconducting channels are competing, so that numerical details decide which one is leading. As the model studied here has so far not been realized experimentally, we refrain from  a classification of the pairings in this two-orbital situation and from elaborating this situation with refined numerics further. However, a general conclusion from this study is that, upon doping, the interaction-driven instabilities in such QBCP systems generically give way to unconventional superconducting phases.

\section{Conclusion}
In summary, we have investigated the weak coupling instabilities of the QBCP in a two-dimensional fermionic spin-$1/2$ system with an angle-resolved fRG calculation. Comparing the leading instability for different ratios of intraorbital and interorbital interactions, we obtained similar results for the continuum model and the model on a checkerboard lattice. For weak interorbital interaction, the system is unstable toward a spin-nematic (SN) phase. For intermediate $U'$, we encounter the time-reversal-symmetry breaking QAH phase, and for a strong interorbital repulsion, the leading tendency is toward the topological nontrivial QSH phase. 
Upon moving the chemical potential away from the band crossing point, the exotic instabilities are replaced by unconventional pairing instabilities at lower critical scale. 

Our results with the QBCP at the chemical potential confirm the conclusion from the mean-field analysis by Sun \textit{et al.} \cite{sun}, mainly obtained for the spinless case. They show that the dominant weak-coupling instabilities of a many-fermion system are not only determined by the shape of the dispersion, but that also the wave-vector-dependent orbital composition of the bands has a decisive impact on the preferred ordering tendency.  This holds as well for linear band crossing points, also known as Dirac points. Here, however, the density of states vanishes at the Fermi level, and nonzero, possibly too large interactions strengths are required to find instabilities.

Further research should address where quadratic and other band crossing points can be found in realistic band structures near the Fermi level in order to investigate their potential instabilities. Bilayer graphene is known to provide two QBCPs at the $K$ and $K'$ points of the Brillouin zone \cite{castroneto}, at least if trigonal warping is ignored. The different spontaneous  quantum Hall states in $N$-layer graphene systems have been classified recently in Ref. \cite{zhangjung}, irrespective of what interactions might be necessary to stabilize these states.  Naively, we suspect however that the bilayer case of two QBCPs with short-range interactions will be dominated by instabilities with the wave-vector connecting the degeneracy points, leading to density wave states, as found in a related approach recently by Vafek \cite{vafek2010}.  Interestingly, for a screened Coulomb interaction with dominant scattering with small wave-vector transfer within the neighborhood of given band crossing point, SN and QAH instabilities were also reported for the bilayer situation \cite{vafek2010,vafekyang2010}. Also, mean-field+fluctuations studies of the bilayer model with long-range Coulomb interaction support the possibility of QAH and QSH states \cite{levitov}. Our study with a single band crossing connects well to these finding and shows that if the two crossing point regions are not at all connected by scattering, even QSH instabilities are possible. Future work should map out the full phase diagram of the bilayer system depending on the screening and also trigonal warping.

This work was supported by the DFG research unit FOR 1162. We thank J. Ortloff, S. Raghu, O. Vafek and S.C. Zhang for discussions.

\bibliographystyle{plaindin}
\bibliography{bib}

%\bibliographystyle{apsrev}
%\bibliography{lowenergyhamiltonians}% Produces the bibliography via BibTeX.

%

 \end{document}